\documentclass[twocolumn,amssymb,amsmath,aps,pra,superscriptaddress,showpacs,showkeys,preprintnumbers,reprint,natbib]{revtex4-1}

\usepackage{graphicx,color}
\usepackage{multirow}
\usepackage{subfig}
\usepackage[normalem]{ulem}

\begin{document}

\title{High-Resolution Laser Spectroscopy of Long-Lived Plutonium Isotopes}

\author{A.~Voss}
\email{anvoss@jyu.fi}
\affiliation{University of Jyv\"askyl\"a, Department of Physics, 40014 Jyv\"askyl\"a, Finland}

\author{V.~Sonnenschein}
\altaffiliation{present address: Nagoya University, Department of Quantum Engineering, Nagoya 464-8603, Japan}
\affiliation{University of Jyv\"askyl\"a, Department of Physics, 40014 Jyv\"askyl\"a, Finland}

\author{P.~Campbell}
\affiliation{The University of Manchester, School of Physics and Astronomy, Manchester M13 9PL, United Kingdom}

\author{B.~Cheal}
\affiliation{University of Liverpool, Oliver Lodge Laboratory, Liverpool L69 7ZE, United Kingdom}

\author{T.~Kron}
\affiliation{Johannes-Gutenberg Universit\"at Mainz, Institut f\"ur Physik, 55128 Mainz, Germany}

\author{I.D.~Moore}
\email{iain.d.moore@jyu.fi}
\affiliation{University of Jyv\"askyl\"a, Department of Physics, 40014 Jyv\"askyl\"a, Finland}

\author{I.~Pohjalainen}
\affiliation{University of Jyv\"askyl\"a, Department of Physics, 40014 Jyv\"askyl\"a, Finland}

\author{S.~Raeder}
\altaffiliation{present address: Helmholtz-Institut Mainz, 55128 Mainz, Germany}
\affiliation{KU Leuven, Instituut voor Kern- en Stralingsfysica, 3001 Leuven, Belgium}

\author{N.~Trautmann}
\affiliation{Johannes-Gutenberg Universit\"at Mainz, Institut f\"ur Kernchemie, 55128 Mainz, Germany}

\author{K.~Wendt}
\affiliation{Johannes-Gutenberg Universit\"at Mainz, Institut f\"ur Physik, 55128 Mainz, Germany}

\date{\today}

\begin{abstract}
Long-lived isotopes of plutonium were studied using two complementary techniques, high-resolution resonance ionisation spectroscopy (HR-RIS) and collinear laser spectroscopy (CLS). Isotope shifts have been measured on the $5f^67s^2\ ^7F_0 \rightarrow 5f^56d^27s\ (J=1)$ and $5f^67s^2\ ^7F_1 \rightarrow 5f^67s7p\ (J=2)$ atomic transitions using the HR-RIS method and the hyperfine factors have been extracted for the odd mass nuclei $^{239,241}$Pu. Collinear laser spectroscopy was performed on the $5f^67s\ ^8F_{1/2} \rightarrow J=1/2\; (27523.61\text{cm}^{-1})$ ionic transition with the hyperfine $A$ factors measured for $^{239}$Pu. Changes in mean-squared charge radii have been extracted and show a good agreement with previous non-optical methods, with an uncertainty improvement by approximately one order of magnitude. Plutonium represents the heaviest element studied to date using collinear laser spectroscopy.
\end{abstract}


\maketitle

\section{Introduction}
Laser spectroscopy is an established technique at radioactive ion beam (RIB) facilities for the study of nuclear shape, size, moments and spins of short-lived radioactive nuclei~\cite{Blaum2013,Campbell2016}. Thus far, the heaviest isotopic chain for which nuclear moments and mean-squared charge radii have been extracted from on-line experiments is that of Ra ($Z=88$)~\cite{Ahmad1988}, above which lie the actinide elements covering a range from Ac ($Z=89$) to Lr ($Z=103$). Such elements are not available at on-line isotope separator facilities and can only be produced via fusion reactions in heavy-ion collisions, transfer reactions using radioactive targets, or, for the study of long-lived isotopes of transuranium elements, bred in sufficient quantities in nuclear reactors and safely transported to facilities equipped for the study of nuclear structure. The combination of low production cross-sections, the lack of stable isotopes and correspondingly only a limited number of reliably determined optical transitions available from literature adds to the challenge of performing laser spectroscopy on these heaviest elements. Due to the scarcity of ground state nuclear structure information in this region of the nuclear chart, efforts are under way to develop suitable techniques which provide the required sensitivity to efficiently make use of the limited quantity of isotopes which can be produced~\cite{Ferrer2013,Backe2015,Ferrer2017}.

Current studies at RIB facilities predominantly use two main techniques of optical spectroscopy. The first, collinear laser spectroscopy, has been applied in a number of variants to the majority of elements, possessing a high resolution which routinely provides measurements of optical frequency splittings to $1-10\text{MHz}$ precision, and a high sensitivity, with minimum fluxes of $\sim 100$ particles per second quoted for systems with hyperfine structure~\cite{Bissell2007}, even lower for even-even isotopes. The second, resonance ionisation spectroscopy (RIS), has been successfully applied directly within the ion source ~\cite{Fedosseev2012} allowing the study of even more exotic nuclei, with minimum half-lives approaching $\sim 1\text{ms}$ and with spectroscopic information extracted from fluxes of below 1 ion per second~\cite{deWitte2007}.

The disadvantage of the in-source RIS method arises from the effect of the different broadening mechanisms which limits the resolution to typically a few GHz. It remains a challenge to analyse lower resolution RIS spectra which exhibit either fully or partially overlapping hyperfine structures and to reliably assign systematic uncertainties to such measurements. Nevertheless, the complementarity of both spectroscopic methods has been demonstrated in recent studies of the nuclear structure of exotic Cu isotopes, where the lower resolution in-source RIS method was often used to greatly reduce the scanning range for high-resolution collinear laser spectroscopy~\cite{Stone2008,Cocolios2009,Cocolios2010,Flanagan2009,Vingerhoets2011,Koster2011}.

In recent years, in-source spectroscopy has improved with the development of advanced cavity designs for pulsed lasers. These new resonators combine the features of high output powers required for the saturation of atomic transitions in the resonant ionisation process, as well as a reduction in the laser linewidth permitting a higher spectroscopic resolution for studies of isotope shifts and hyperfine structures. A ring design of Ti:Sapphire laser cavity developed at Mainz University resulted in output powers of up to $1\text{W}$ in single-mode operation, with a laser linewidth of below $50\text{MHz}$~\cite{deGroote2015}. Injection-locked systems on the basis of a similar ring cavity Ti:Sapphire laser have demonstrated linewidths of pulsed high-repetition-rate radiation of below $20\text{MHz}$, while maintaining an impressive output power in the range of a few Watts~\cite{Kessler2008,Sonnenschein2015}. 

In this work, a new programme of heavy element research at the \textsc{Igisol} facility in the Accelerator Laboratory of the University of Jyv\"askyl\"a was initiated in collaboration with the Institut f\"ur Physik, Johannes-Gutenberg Universit\"at, Mainz. Plutonium ($Z=94$) was chosen to be a suitable candidate having a number of long-lived isotopes, $^{238-244}$Pu, for which sufficiently large sample sizes~(ng) have been supplied by the Institut f\"ur Kernchemie, Mainz, for studies both in Mainz and at Jyv\"askyl\"a. Earlier optical emission studies of Pu include a measurement of the isotope shift for $^{239-240}$Pu with modest resolution in a number of atomic levels and transition lines~\cite{ActinideTables}, whereas for a larger set of isotopes, $^{238-242}$Pu, the shift was determined with a precision of approximately $100\text{MHz}$ in selected levels in neutral Pu~\cite{LandoltBornstein}. Motivated by trace analysis applications \cite{Trautmann2004, Raeder2012}, resonance ionisation has been applied to quantify the plutonium amount in environmental samples. To enable isotope selectivity in these studies, resonance ionisation spectroscopy (RIS) was used in combination with a time-of-flight mass spectrometer to resolve the isotope shifts in $^{238-242,244}$Pu~\cite{Gruning2001} to a precision of about $600\text{MHz}$, with later refinements using narrow linewidth continuous-wave (CW) lasers resulting in a precision of $15-30\text{MHz}$ for $^{239,240,242,244}$Pu~\cite{Kunz2004}.

The focus of the current article is the first comparison of the experimental techniques of collinear laser spectroscopy and high-resolution RIS in the actinide region. A measurement of optical isotope shifts in Pu has been performed on three transitions, two atomic transitions using an injection-locked pulsed laser system at Mainz and an ionic transition via collinear laser spectroscopy in Jyv\"askyl\"a. The King plot method~\cite{King1984} is used to extract changes in mean-squared charge radii in order to assess the accuracy of both techniques. Establishing a good agreement is of importance such that either method can be used in the future to explore the properties of heavier actinides or lighter refractory elements which have thus far been challenging to access.

The article is structured as follows: in section~\ref{sec:exp} details of the two techniques are presented. The results and data analysis for the two experiments are presented in section~\ref{sec:data}, with the extraction of changes in mean-squared charge radii in section~\ref{sec:king}. Section~\ref{sec:sys_errors} contains a discussion of the systematic errors assigned to both techniques. The final conclusions are drawn in section~\ref{sec:conc}.

\section{Experimental Technique}
\label{sec:exp}

As this work utilises two very different laser spectroscopic techniques, resonance ionisation spectroscopy using pulsed lasers and collinear laser spectroscopy using a continuous-wave laser, both techniques are described separately.

\subsection{High-Resolution Resonance Ionisation Spectroscopy}

\begin{figure}
\subfloat[\textsc{Mabu}]{\includegraphics[width=\columnwidth]{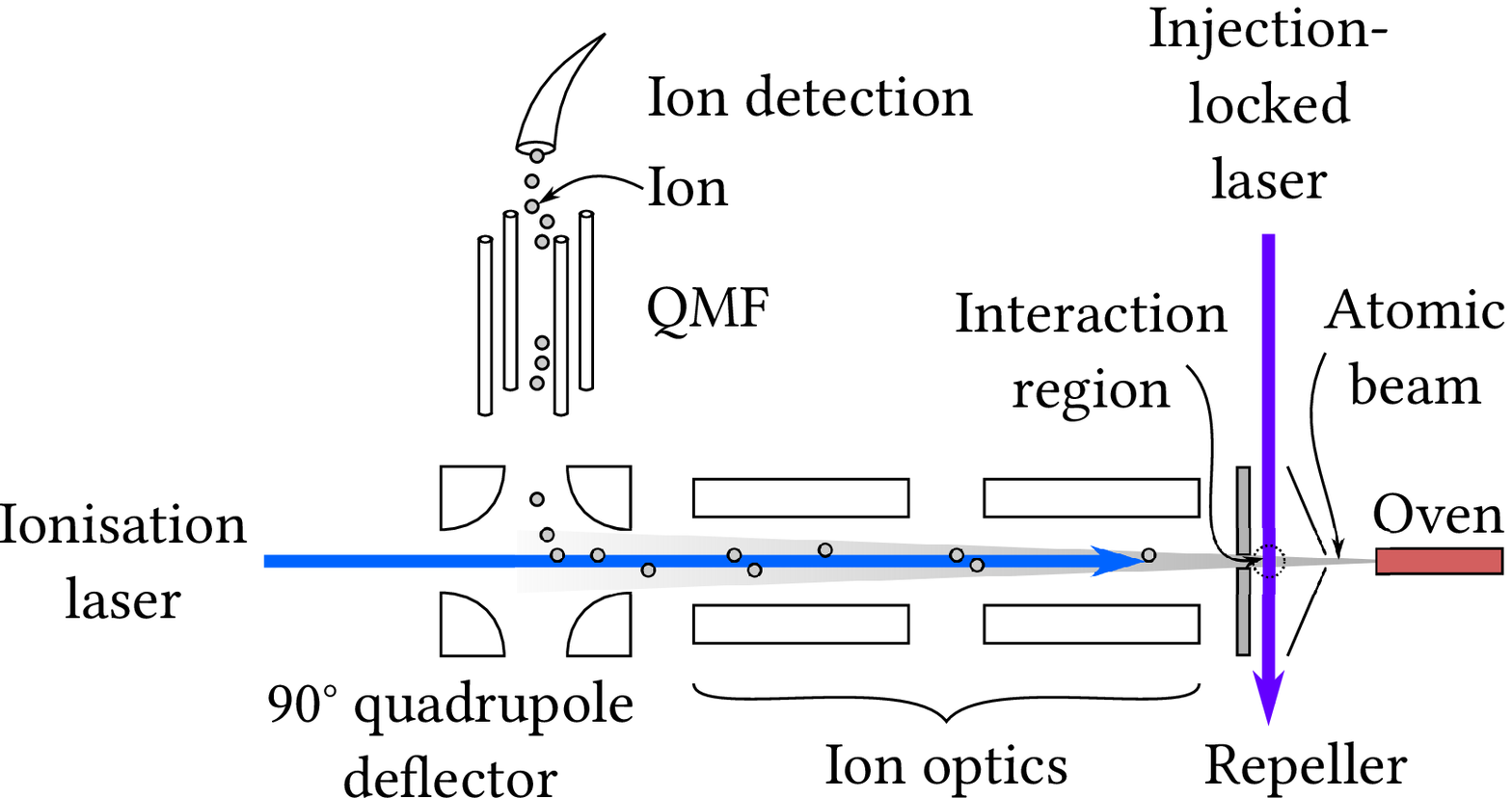}\label{fig:mabu}}\\
\subfloat[HR-RIS schemes]{\includegraphics[width=\columnwidth]{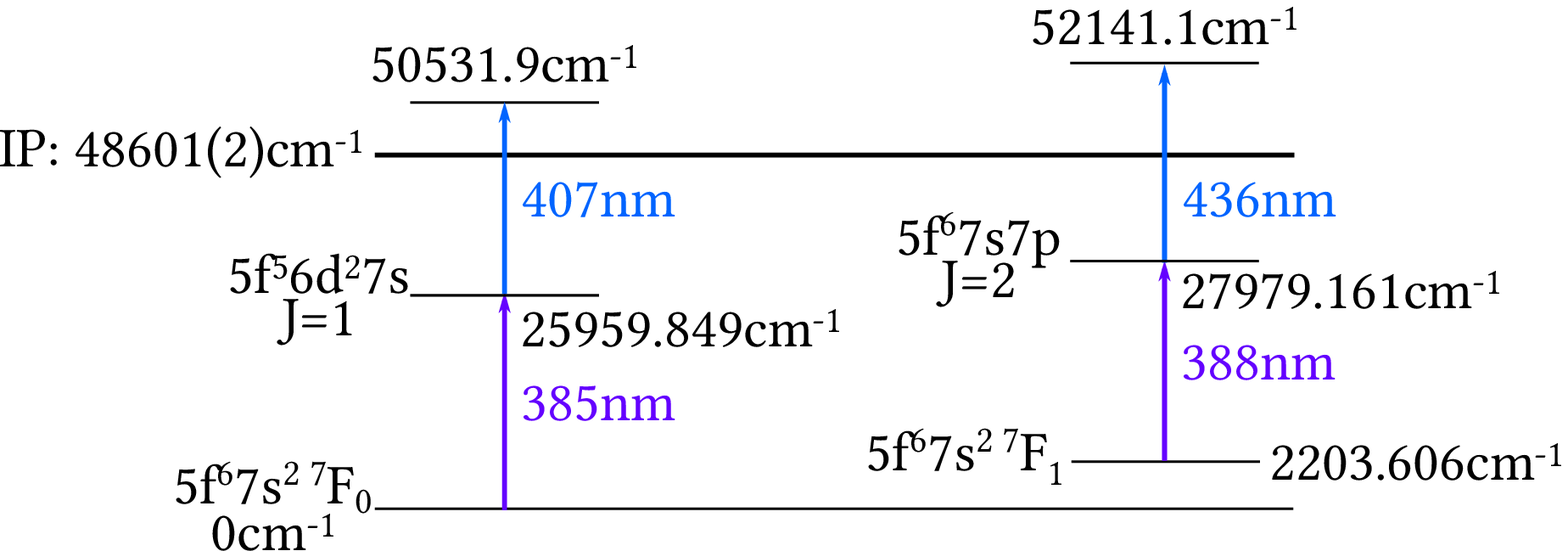}\label{fig:scheme}}
\caption{(Colour online) Experimental setup at Mainz. (a) Schematic depiction of the \textsc{Mabu} QMF system; see text for details. (b) Optical scheme used for HR-RIS. The wavelength of the first step was scanned. Data for level configurations and energies taken from~\cite{ActinideTables,Koehler1997}.}
\end{figure}

High-resolution resonance ionisation spectroscopy (HR-RIS) was performed using a Ti:sapphire laser system composed of one conventional $10\text{kHz}$ high repetition rate laser together with a dedicated injection-locked Ti:Sapphire laser~\cite{Sonnenschein2015} which ensured a specifically narrow bandwidth. These were operated at the Mainz Atomic Beam Unit (\textsc{Mabu}) which comprises a quadrupole mass filter (QMF), as schematically depicted in Fig.~\ref{fig:mabu}~\cite{Rossnagel2012}. A well-collimated atomic beam was formed by resistively heating a graphite oven to approximately $1300\text{K}$. In order to reduce the large Doppler broadening due to the thermal velocity distribution of the atomic ensemble expected in co-/counter-propagating laser irradiation, the laser beam from the injection-locked Ti:Sapphire laser was introduced in a perpendicular geometry to the effusing atomic beam and expanded to generate a uniform intensity distribution within the interaction volume. The ionisation laser was introduced counter-propagating to the atomic beam and focused to a spot size comparable to the oven dimension with an inner diameter of $2\text{mm}$. Following resonant ionisation, the ion beam was shaped and deflected by $90^\circ$ in a quadrupole deflector to enter the QMF (mass resolving power $M/\Delta M \sim 200$) and subsequently detected using a channeltron electron multiplier operating in single-ion counting mode. Due to the pulsed nature of the lasers, a time-gating method was employed to minimise the background from surface ions.

The ring cavity Ti:Sapphire laser was injection-locked to an external cavity diode laser (\textsc{Ecdl}) via a single-mode optical fibre providing $5-20\text{mW}$ seed input. The spectral linewidth of the Ti:Sapphire laser was analysed with a commercial scanning Fabry-P\'erot interferometer (FPI) with a free spectral range (FSR) of $~300\text{MHz}$, resulting in a measured linewidth of $13.4(8)\text{MHz}$. This may be compared with the measured linewidth of the master (\textsc{Ecdl}) laser of $10.1(2)\text{MHz}$. The \textsc{Ecdl} was stabilised via a quadrature interferometer (\textsc{iScan}, \textsc{Tem Messtechnik}) for fast frequency control in combination with fringe-offset locking for long-term stability~\cite{Fischbach2012,*Hakimi2013}. By locking to a confocal FPI ($\text{FSR}=299.782(5)\text{MHz}$) and using a frequency-stabilised HeNe laser as reference, a frequency calibration of better than $1\text{MHz}$ could be attained. The frequency of the injection-locked laser was scanned in a stepwise manner by driving the \textsc{Ecdl} to fixed setpoints. Data acquisition of the ion signal took place whenever the \textsc{Ecdl} laser frequency was within a $\pm5\text{MHz}$ locking interval of the setpoint.

Spectroscopy on neutral Pu was performed on two different atomic transitions illustrated in Fig.~\ref{fig:scheme}. The first transition at $385.210\text{nm}$ proceeds from the atomic ground state ($J=0$) to an excited state at $25959.849\text{cm}^{-1}$ ($J=1$), whereas the second transition at $387.965\text{nm}$ proceeds from a thermally populated state at $2203.606\text{cm}^{-1}$ ($J=1$) to a level at $27929.161\text{cm}^{-1}$ ($J=2$). At an oven temperature of $\sim 1300\text{K}$ the metastable state is expected to have a 20\% population with respect to the ground state. The wavelengths were obtained by single-pass frequency doubling of the laser radiation from the injection-locked Ti:Sapphire laser using a $\beta$-barium borate (BBO) non-linear crystal. The ionisation step for both transitions proceeded via auto-ionising states above the ionisation potential (IP) and was provided by an intra-cavity frequency doubled broadband Ti:Sapphire laser with a fundamental linewidth of approximately $4-5\text{GHz}$. Typical laser powers available for the ionisation step were up to $1\text{W}$. Lower laser powers of the order of $2-10\text{mW}$ were used for the spectroscopy step to minimise saturation broadening.

\subsection{Collinear Laser Spectroscopy}

Collinear laser spectroscopy was performed at the \textsc{Igisol} facility of the Accelerator Laboratory at the University of Jyv\"askyl\"a. Samples containing Pu isotopes ($^{238-242,244}$Pu) were electrolytically deposited onto a tantalum substrate which was electrothermally heated inside a gas-cell filled with helium. The Pu atoms were selectively ionised via two-step resonant laser ionisation utilising two intra-cavity frequency doubled, broadband Ti:Sapphire lasers operating at a repetition rate of $10\text{kHz}$ and a linewidth of $\sim100\text{GHz}$. Further details concerning the gas-cell designed for such heavy element studies as well as the in-gas-cell resonant laser ionisation process have been published elsewhere~\cite{Pohjalainen2016}.

\begin{figure*}
\includegraphics[width=\textwidth]{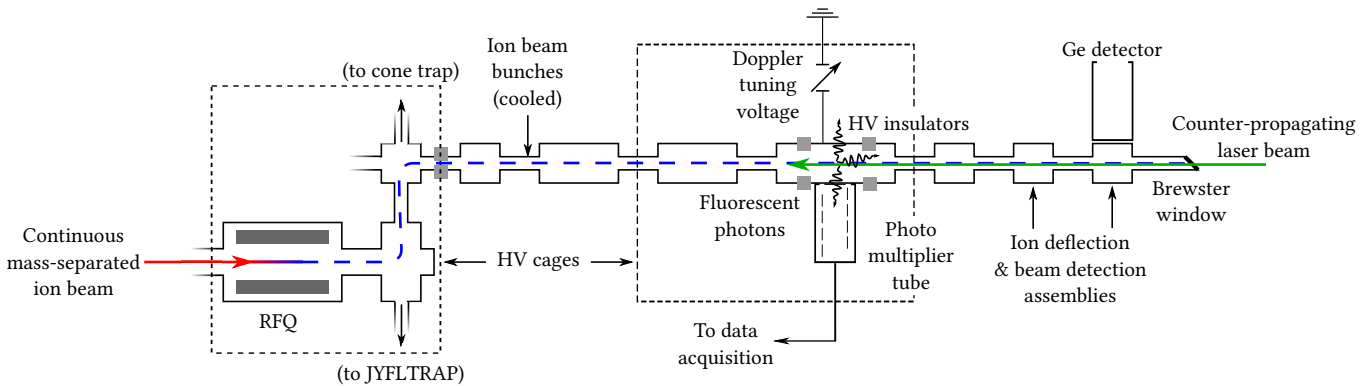}
\caption{(Colour online) Schematic diagram of the collinear laser spectroscopy beamline at \textsc{Jyfl}. The bunched and cooled ion beam is overlapped with a counter-propagating, narrow linewidth laser beam. A Ge detector may be used for the detection of gamma rays emitted following the decay of a radioactive species and is often used for ion beam identification during initial tuning.}
\label{fig:laserline}
\end{figure*}

The ions were extracted from the gas-cell via gas flow, guided through a sextupole ion guide (SPIG)~\cite{Karvonen2008} and accelerated to $30\text{keV}$ towards a mass separator with a typical resolving power $M/\Delta M \sim 350$. Following mass separation, a continuous ion beam of a single $A/q$ was injected into a gas-filled radio-frequency Paul trap (RFQ) for cooling and bunching~\cite{Nieminen2001}. The use of an RFQ in conjunction with collinear laser spectroscopy was pioneered at the \textsc{Igisol} facility in order to suppress the laser-scattered background by gating the data acquisition with respect to the arrival of an ion bunch at the light collection region~\cite{Nieminen2002}. The bunched ion beam was overlapped in a collinear geometry with a counter-propagating laser beam. A scanning voltage applied to the light collection region Doppler shifted the ions into resonance with the laser light and the resulting fluorescent photons were imaged and detected on a photomultiplier tube. A schematic diagram of the collinear laser spectroscopy beamline is given in Fig.~\ref{fig:laserline}. In this work, the tantalum filaments were heated to a temperature of approximately $1300-1500\text{K}$, depending on the abundance of the isotope of interest to be evaporated from the filament, such that a Pu$^{+}$ ion yield of approximately $30,000/\text{s}$ were detected on a set of microchannel plates at the end of the collinear laser spectroscopy beamline.

Laser spectroscopy was performed from the PuII ionic ground state on the $5f^67s\ ^8F_{1/2} \rightarrow J=1/2\; (27523.61\text{cm}^{-1})$ transition~\cite{ActinideTables} at $363.324\text{nm}$. The laser light was generated by a \textsc{SpectraPhysics 380D} dye laser operating with Pyridine2 dye pumped by a \textsc{Coherent Verdi V5} diode-pumped solid-state (DPSS) laser at $532\text{nm}$. Long-term frequency stabilisation was achieved employing a ``top-of-fringe'' locking to an iodine absorption line with a $3\text{MHz}$ accuracy. Intra-cavity frequency doubling using a BBO crystal allowed the generation of up to $0.35\text{mW}$ UV light which was injected into the beamline through a Brewster window.

\section{Data Analysis and Results}
\label{sec:data}

\subsection{Resonance Ionisation Spectroscopy}

\begin{figure}
\subfloat[HR-RIS, $385.210\text{nm}$]{\includegraphics[width=\columnwidth]{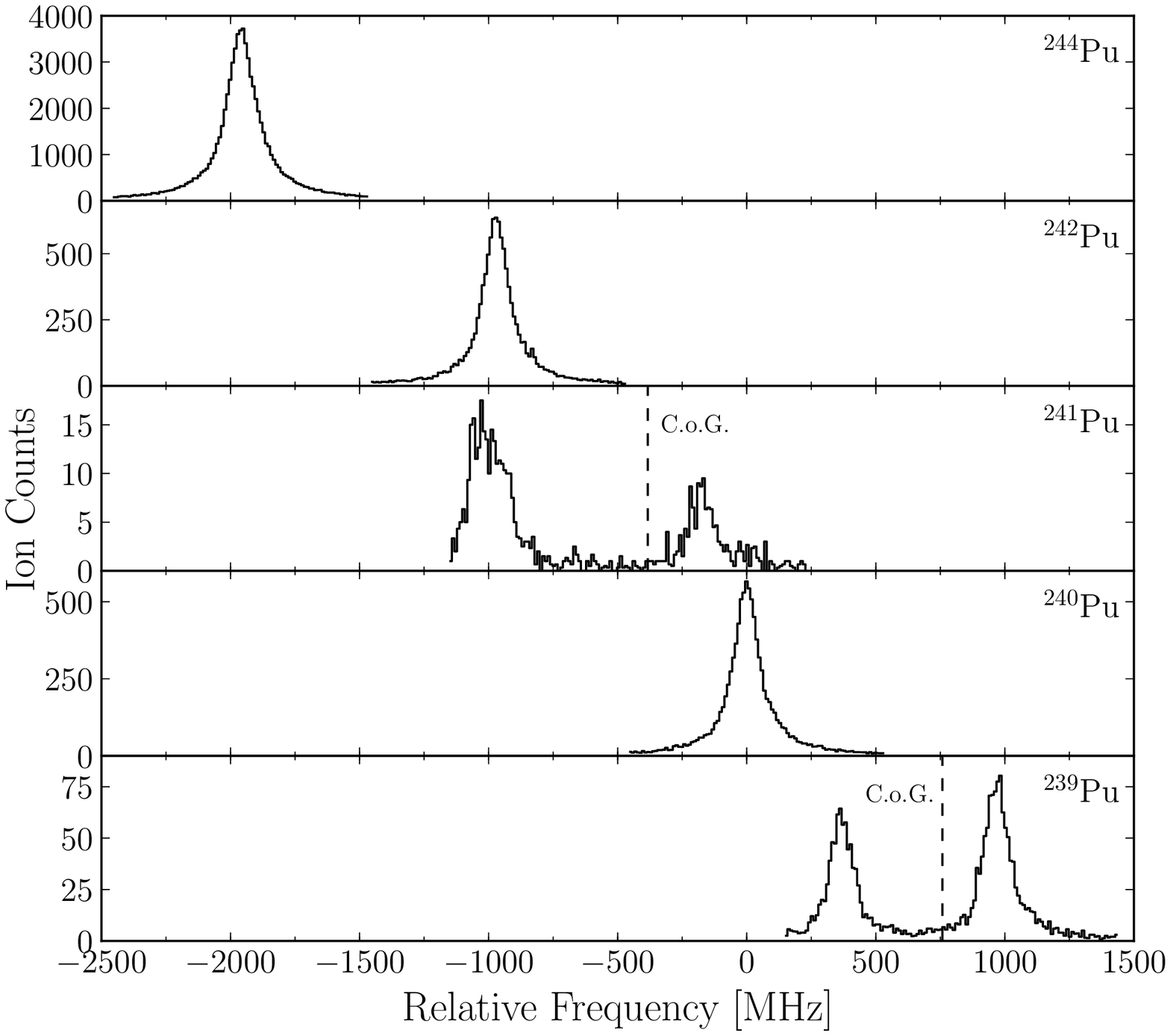}\label{fig:RIS_A}}\\
\subfloat[HR-RIS, $387.965\text{nm}$]{\includegraphics[width=\columnwidth]{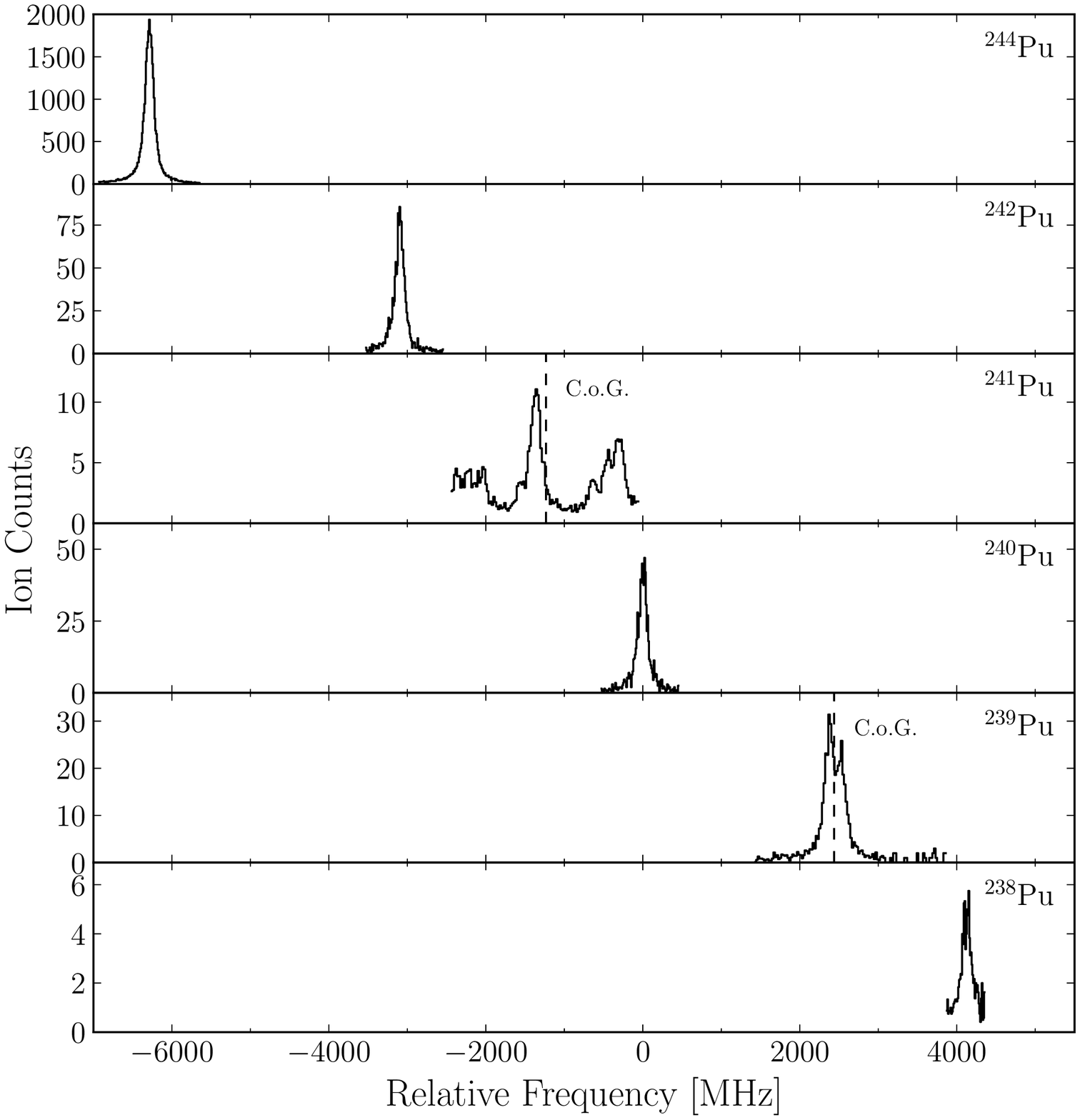}\label{fig:RIS_B}}
\caption{(Colour online) High-resolution resonance ionisation spectra using the injection-locked Ti:Sapphire laser for the two transitions investigated.}
\label{fig:RIS-spectra}
\end{figure}

The resonance ionisation spectra obtained using the two transitions on $^{238-242,244}$Pu atoms are shown in Fig.~\ref{fig:RIS-spectra}. A frequency jitter of the \textsc{Ecdl} of typically $5-10\text{MHz}$ and the nature of the data acquisition selection ($\pm5\text{MHz}$ around the setpoint for the fundamental light) led to small shifts of the resonance positions between the two scan directions. In order to consider this effect, the data from scans with increasing and decreasing frequency were summed for the analysis process. The fit results for the hyperfine coefficients and isotope shifts with respect to the reference isotope $^{240}$Pu are given in Table~\ref{tab:RIS-results}.

\begin{table*}
\caption{Overview of the extracted hyperfine parameters and isotope shifts $\delta\nu^{240,A}$ (all in MHz) for the two transitions studied in this work. Empty fields indicate that the corresponding parameter cannot be determined due to the nuclear spin $I$. Statistical uncertainties arising from fits to the data are denoted by round brackets whereas systematic uncertainties are given in square brackets (see section~\ref{sec:sys_errors_RIS}). The term \{ref\} for $^{240}$Pu indicates the reference nature of this isotope. Uncertainties arising from the determination of its resonance centroid amount to $3.0\text{MHz}$ and have been folded into the isotope shifts for the other isotopes.}
\label{tab:RIS-results}
\begin{tabular}{c c | r@{.}l r@{.}l r@{.}l | r@{.}l r@{.}l r@{.}l r@{.}l r@{.}l}
\hline\hline
        &     & \multicolumn{6}{c|}{$385.210\text{nm}$}                                                                                & \multicolumn{10}{c}{$387.965\text{nm}$}\\
Isotope & $I$ & \multicolumn{2}{c}{$\delta\nu^{240,A}$} & \multicolumn{2}{c}{$A_\text{upper}$} & \multicolumn{2}{c|}{$B_\text{upper}$}     & \multicolumn{2}{c}{$\delta\nu^{240,A}$} & \multicolumn{2}{c}{$A_\text{lower}$} & \multicolumn{2}{c}{$A_\text{upper}$}          & \multicolumn{2}{c}{$B_\text{lower}$} & \multicolumn{2}{c}{$B_\text{upper}$} \\
\hline
$^{244}$Pu & $0$   & $-$1955&2(55)[80]        & \multicolumn{2}{c}{} & \multicolumn{2}{c|}{} & $-$6288&1(60)[80]   & \multicolumn{2}{c}{}           & \multicolumn{2}{c}{} & \multicolumn{2}{c}{} & \multicolumn{2}{c}{} \\
$^{242}$Pu & $0$   &  $-$969&4(54)[80]        & \multicolumn{2}{c}{} & \multicolumn{2}{c|}{} & $-$3099&3(90)[80]   & \multicolumn{2}{c}{}           & \multicolumn{2}{c}{} & \multicolumn{2}{c}{} & \multicolumn{2}{c}{} \\
$^{241}$Pu & $5/2$ &  $-$379&5(117)[80]$^\ast$ & $-$278&1(56)         & $+$100&8(180)         & $-$1228&3(120)[80]  & $-$2&2(48)                     & $+$37&7(34)          & $+$1058&6(82)        & $-$175&4(143)        \\
$^{240}$Pu & $0$   &       0&0\{ref\}         & \multicolumn{2}{c}{} & \multicolumn{2}{c|}{} &       0&0\{ref\}    & \multicolumn{2}{c}{}           & \multicolumn{2}{c}{} & \multicolumn{2}{c}{} & \multicolumn{2}{c}{} \\
$^{239}$Pu & $1/2$ &  $+$757&8(60)[80]        & $+$402&3(32)         & \multicolumn{2}{c|}{} & $+$2438&7(120)[80]  & \multicolumn{2}{c}{0$^\ddagger$} & $-$62&6(33)          & \multicolumn{2}{c}{} & \multicolumn{2}{c}{} \\
$^{238}$Pu & $0$   &  \multicolumn{6}{c|}{not studied using this transition}                 & $+$4126&7(130)[80]  & \multicolumn{2}{c}{}           & \multicolumn{2}{c}{} & \multicolumn{2}{c}{} & \multicolumn{2}{c}{} \\ 
\hline\hline
\multicolumn{18}{l}{$\ast$ Not an independent measurement. See text for details.}\\
\multicolumn{18}{l}{$\ddagger$ Fixed parameter. See text for details.}\\
\end{tabular}
\end{table*}

\begin{figure}
\subfloat[Hyperfine level diagram]{\includegraphics[width=.5\columnwidth]{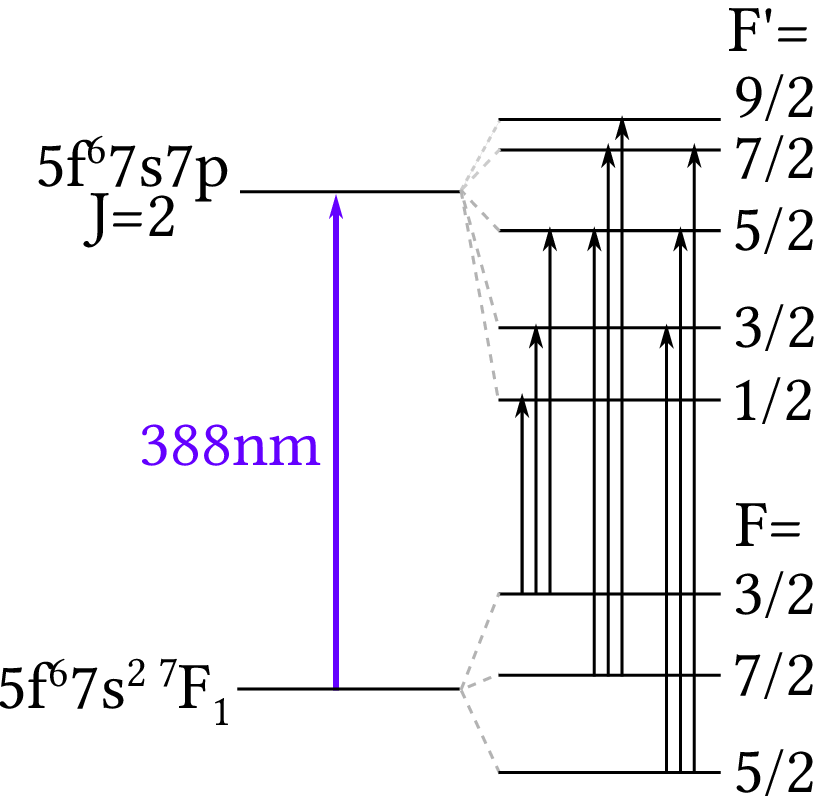}\label{fig:hfs}}\\
\subfloat[Hyperfine spectrum]{\includegraphics[width=\columnwidth]{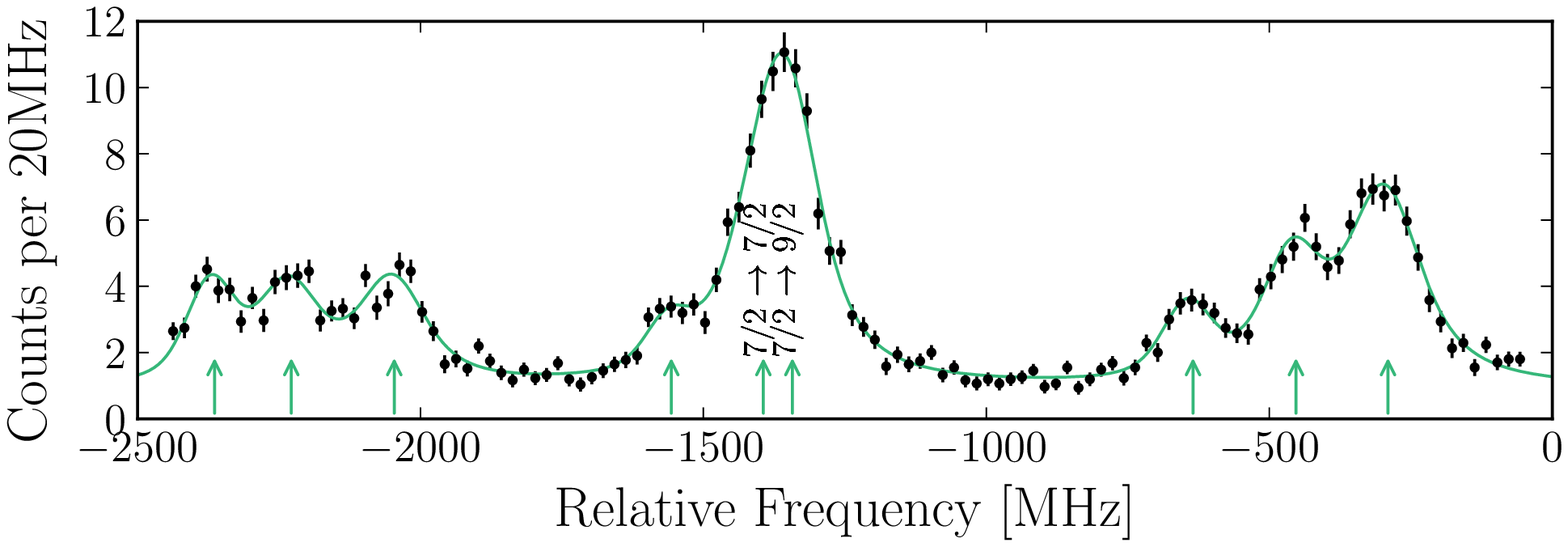}\label{fig:RIS-fit}}
\caption{(Colour online) Expanded hyperfine level diagram and example fit for the $5f^67s^2\ ^7F_1 \rightarrow 5f^67s7p\ (J=2)$ ($387.965\text{nm}$) HR-RIS transitions for $^{241}$Pu ($I=5/2$). Allowed transitions are indicated by arrows in the level diagram. A hyperfine structure fit with free intensities is included as the solid line to highlight the quality of the fit with respect to the small errorbars on the data points. The locations of the hyperfine transitions are indicated by arrows with those labelled corresponding to the unresolved transitions.}
\end{figure}

The data were fitted with a Voigt lineshape whereby the Gaussian component was considerably smaller than the Lorentzian component. The oven temperature was adjusted to optimise the release of the isotope of interest, which in turn may provide a different Gaussian contribution to the linewidth of each isotope. Free parameters in the fit were the background, the lineshape (FWHM, Gaussian and Lorentzian contributions), the peak intensities and its centroid as well as hyperfine parameters where applicable. For each isotope, one lineshape was assumed for all resonances. The FWHM of the resonances varied from $100-150\text{MHz}$ of which approximately $20\%$ was attributable to the Gaussian component. With respect to the hyperfine parameters, some additional constraints were included in the fitting procedure as outlined in the following. An example fit of the $^{241}$Pu spectrum for the $387.965\text{nm}$ atomic transition is given in figure~\ref{fig:RIS-fit} using free intensities to show the close correspondence of the fitted spectrum with regards to the experimental errorbars.

In the metastable transition at $387.965\text{nm}$, the peaks for the $F=7/2 \rightarrow F'=7/2$ and $F=7/2 \rightarrow F'=9/2$ hyperfine transitions (see Figure~\ref{fig:hfs} for an expanded level scheme) overlap closely in the $^{241}$Pu hyperfine spectrum. The relative intensities for these peaks were therefore fixed to the theoretical value of the corresponding Racah coefficients in the fit. Since the hyperfine $A$ coefficient of the lower state is close to and consistent with zero, as determined from the $^{241}$Pu spectrum, it was kept fixed at zero for the evaluation of the collapsed hyperfine structure in $^{239}$Pu.

The atomic ground state exhibits no splitting due to a spin $J=0$. Additionally, $^{238}$Pu was not observed using the $385.210\text{nm}$ transition which may indicate that despite a higher thermal population in the atomic ground state, the ionisation scheme starting from the metastable state was more efficient. As the \textsc{Mabu} was optimised for high transmission, the obtained mass resolving power of the QMF was not sufficient to fully discriminate between the Pu isotopes and thus the more abundant $^{242}$Pu was also present in scans of the hyperfine structure of $^{241}$Pu. As the hyperfine component of lowest frequency overlaps with the resonance of $^{242}$Pu, the extraction of the isotope shift was more demanding. It was also noticed during the data analysis, that the third component of the $^{241}$Pu hyperfine spectrum of highest frequency was missed during the laser scans. By using the ratio of field shifts between the two transitions for the even isotopes, the hyperfine structure centroid of $^{241}$Pu for the ground state transition can be estimated as
\begin{equation}
\delta \nu_{385}^{240,241} = \frac {F_{385}}{F_{388}} \times \delta \nu_{388}^{240,241}\;\text{,}
\end{equation}
where $F_{385}/F_{388}=0.309(7)$ from a King plot. Included in the fitting of $^{241}$Pu were the contributions from the abundant $^{242}$Pu mass peak as well as a weaker $^{240}$Pu component using the independently determined centroid positions.

\subsection{Collinear Laser Spectroscopy}

As the iodine absorption line at $\sim727\text{nm}$ used to stabilise the dye laser is untabulated, the laser frequency had to be calculated from the resonance spectra of $^{240}$Pu wherefore the transition frequency $\nu_\text{tran}$ is known~\cite{ActinideTables}. This was accomplished by fitting the raw data of fluorescent photons versus DAQ channel number. The resonance channel number $x$ and the non-relativistic equation for a frequency offset $\Delta\nu$ for counter-propagating laser beams,
\begin{subequations}
\begin{align}
\Delta\nu &= \nu_\text{laser}\left(1+\frac{v}{c}\right)-\nu_\text{tran}\\
          &= \nu_\text{laser}\left(1+\sqrt{\frac{2[eV_\text{RFQ}-(mx+b)]}{m_\text{ion}c^2}}\right)-\nu_\text{tran},
\end{align}
\end{subequations}
where $V_\text{RFQ}$ is the bias voltage of the RFQ and $m$, $b$ are the slope and intercept calibration parameters of the scanning voltage, was used. For the reference isotope $^{240}$Pu, $\Delta\nu=0$ on resonance.

The collinear laser spectroscopic work was performed on singly-ionised species and thus a correction to the atomic masses $m_\text{atom}$ from~\cite{Wang2012} has been included to account for the mass difference due to the missing electron. Here,
\begin{equation}
m_\text{ion} = m_\text{atom} - m_\text{electron} + m_\text{IP},\label{eq:m_ion}
\end{equation}
where $m_\text{electron}$ was taken from~\cite{codata2014} and $m_\text{IP}$ represents the ionisation potential converted into mass units from~\cite{Koehler1997}. The effect of the latter is negligible, however, has been included in equation~\eqref{eq:m_ion} for completeness. The calculated ionic masses used for evaluation of the optical spectra are tabulated in Table~\ref{tab:ion_masses}.

The corresponding wavenumber of the ring dye laser was determined to be $\bar{\lambda}_\text{laser} = 13754.625\text{cm}^{-1}$ from the weighted mean of all $^{240}$Pu scans taking into account the frequency doubling. Any statistical errors related to the fits of the optical spectrum do not contribute at this scale. The fractional systematic uncertainty of the voltages due to the readback of the RFQ and scanning voltage of the light collection region corresponds to $0.1\%$~\cite{Charlwood2009}. The effect of a systematic uncertainty arising due to the voltage readback is considered to be $\sim0.0035\text{cm}^{-1}$. An uncertainty for the transition wavenumber $\bar{\lambda}_\text{transition}$ is not given in the literature~\cite{ActinideTables}, however, is assumed to be $0.01\text{cm}^{-1}$. The uncertainty of $\bar{\lambda}_\text{laser}$ is dominated by the uncertainty of $\bar{\lambda}_\text{transition}$. An overall systematic uncertainty of $0.01\text{cm}^{-1}$ for $\bar{\lambda}_\text{laser}$ is therefore assumed.

\begin{table}
\caption{Comparison of the atomic masses from~\cite{codata2014} and the ionic masses used in this work.}
\label{tab:ion_masses}
\begin{tabular}{c r@{.}l r@{.}l}
\hline\hline
$A$ & \multicolumn{2}{c}{$m_\text{atom}$ [amu]} & \multicolumn{2}{c}{$m_\text{ion}$ [amu]} \\
\hline
244 & 244&06420526(557) & 244&06365669(557)\\
242 & 242&05874281(196) & 242&05819424(196)\\
240 & 240&05381375(192) & 240&05326518(192)\\
239 & 239&05216359(192) & 239&05161502(192)\\
\hline\hline
\end{tabular}
\end{table}

\begin{figure}
\includegraphics[width=\columnwidth]{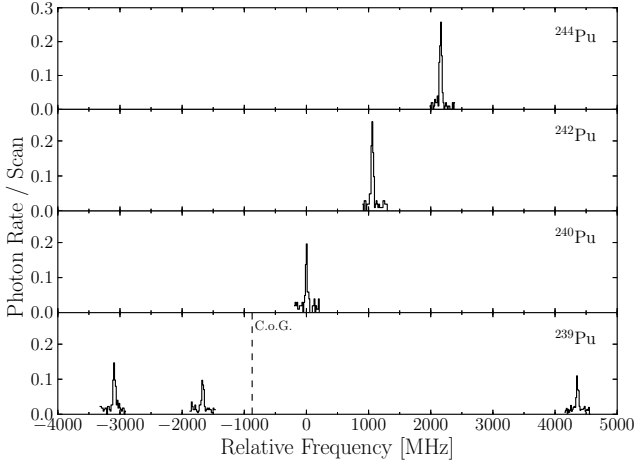}
\caption{(Colour online) Optical fluorescence spectra for $^{244,242,240,239}$Pu$^{+}$ isotopes determined by collinear laser spectroscopy on the $^8F_{1/2} \rightarrow J=1/2$ ($363.324\text{nm}$) ionic transition. The centre of gravity (C.o.G.) in $^{239}$Pu is marked as a vertical dashed line.}
\label{fig:CLS-spectra}
\end{figure}

\begin{figure}
\subfloat[Hyperfine level diagram]{\includegraphics[width=.5\columnwidth]{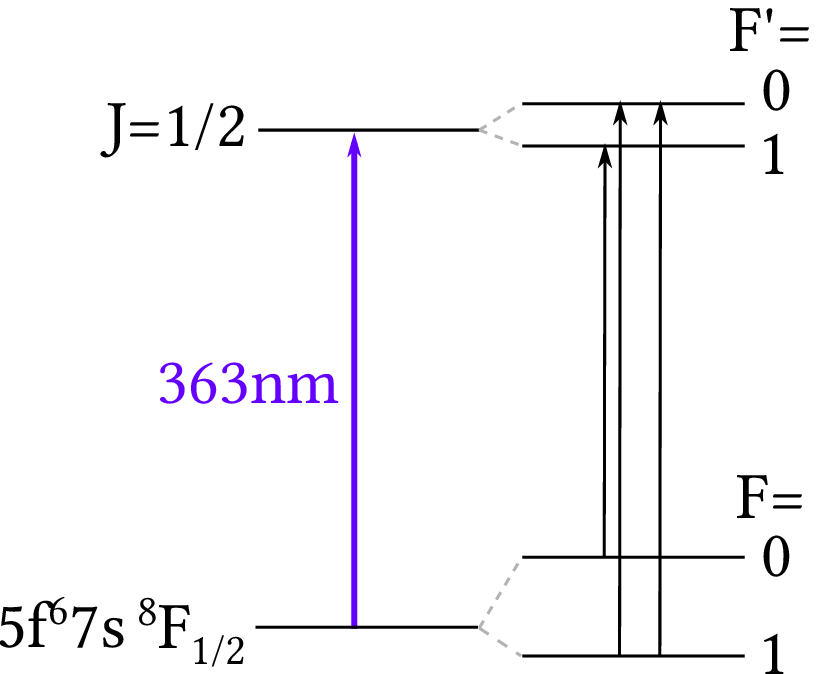}}\\
\subfloat[Hyperfine spectrum]{\includegraphics[width=\columnwidth]{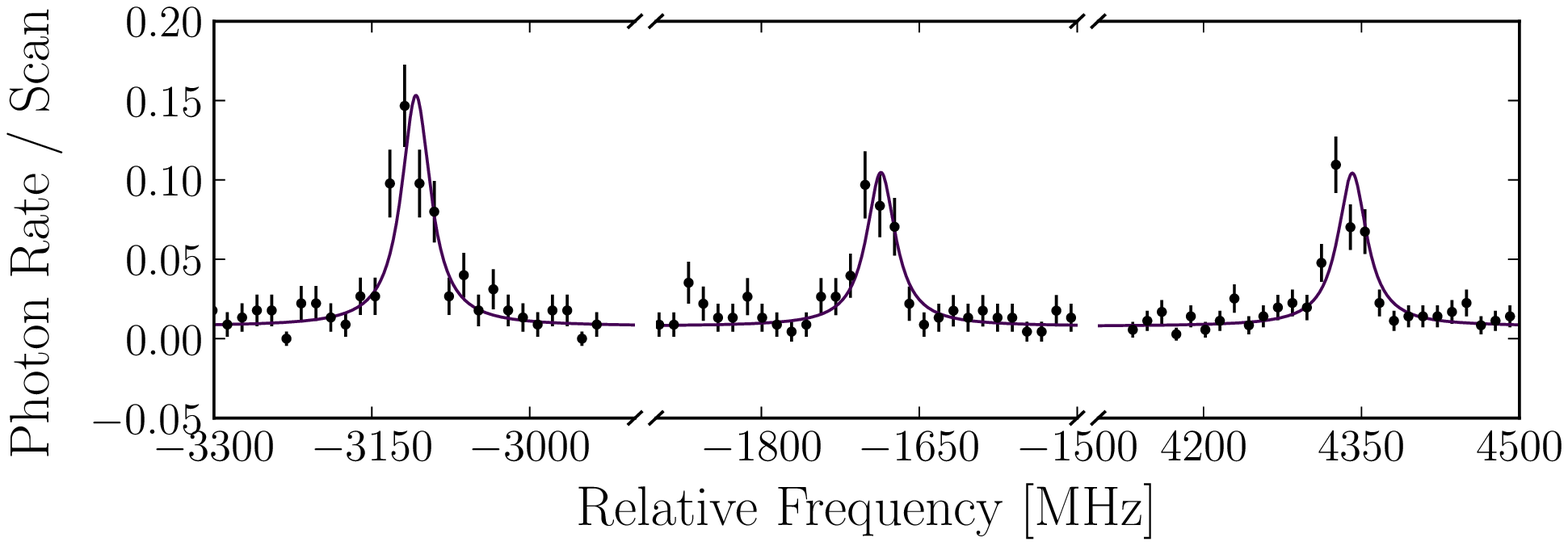}}
\caption{(Colour online) Hyperfine level diagram and example fit for the $5f^67s\ ^8F_{1/2} \rightarrow J=1/2$ ($363.324\text{nm}$) CLS transition for $^{239}$Pu$^+$ ($I=1/2$). A hyperfine structure fit with free intensities is shown as the solid line together with statistical error bars on the data points.}
\label{fig:CLS-fit}
\end{figure}

The optical fluorescence spectra of singly-charged $^{244,242,240,239}$Pu ions measured in this work are shown in Fig.~\ref{fig:CLS-spectra}. A purely Lorentzian lineshape was used in the data analysis reflecting a zero Gaussian contribution to the spectra. This can be expected when using cooled ion beams in which the energy spread is typically $<0.6 \text{eV}$~\cite{Campbell2002}. The fit parameters included the background, the FWHM of the resonance, the centroid and the intensity of the resonance for the $I=0$ isotopes. The FWHM of the resonances was $\sim30\text{MHz}$ for all isotopes. For isotopes with a non-zero nuclear spin, the hyperfine $A$ coefficients of the atomic ground and excited states were allowed to vary and the FWHM of all resonances was constrained to be common. An example of such a fit is shown in figure~\ref{fig:CLS-fit}. The hyperfine structure of $^{239}$Pu was fitted with free intensities with the relative intensities from the best fit parameters corresponding to those from weak field coupling estimates (``Racah intensities''). Due to the choice of optical transition, $J=1/2 \rightarrow J'=1/2$, there is no sensitivity to the electric quadrupole moments. The extracted parameters and isotope shifts with respect to $^{240}$Pu are summarised in Table~\ref{tab:CLS-results}. 

\begin{table}
\caption{Summary of the extracted hyperfine $A$ parameters and isotope shifts $\delta\nu^{240,A}$ (all in MHz) for the collinear laser spectroscopy work on the $363\text{nm}$ ionic transition. Statistical uncertainties arising from hyperfine structure fits to the data are denoted by round brackets whereas systematic uncertainties (see section~\ref{sec:sys_errors_CLS}) due to the conversion from scanning voltages into frequencies are given in square brackets.}
\label{tab:CLS-results}
\begin{tabular}{c c r@{.}l r@{.}l r@{.}l}
\hline\hline
Isotope & $I$ & \multicolumn{2}{c}{$\delta\nu^{240,A}$} & \multicolumn{2}{c}{$A_\text{lower}$} & \multicolumn{2}{c}{$A_\text{upper}$} \\
\hline
$^{244}$Pu & $0$   & $+$2160&6(48)[25] & \multicolumn{2}{c}{} & \multicolumn{2}{c}{} \\
$^{242}$Pu & $0$   & $+$1056&3(42)[12] & \multicolumn{2}{c}{} & \multicolumn{2}{c}{} \\
$^{240}$Pu & $0$   &       0&0\{ref\}  & \multicolumn{2}{c}{} & \multicolumn{2}{c}{} \\
$^{239}$Pu & $1/2$ &  $-$872&7(55)[9]  & $+$7445&5(32)        & $-$1421&0(37)        \\
\hline\hline
\end{tabular}
\end{table}

\section{Comparison of techniques using the King plot method}
\label{sec:king}

Information on the changes in mean-squared charge radii between nuclei with atomic masses $A$ and $A^{'}$ may be extracted from optical isotope shifts, $\delta\nu$, as
\begin{align}
\delta\nu^{A',A} &= \nu^{A} - \nu^{A'}\\
                &= \left(\frac 1 {m_{A'}} - \frac 1 {m_A}\right) M + F K(Z) \delta \langle r^2 \rangle^{A',A}. \label{eq:IS}
\end{align}
Here, $M$ and $F$ are the transition-dependent atomic factors for the mass and field shift, respectively. $K(Z)$ is an element-dependent factor to correct for higher order (Seltzer) moments which contribute a few percent in heavier nuclei ~\cite{Seltzer1969,Torbohm1985}. The atomic factors are to be determined either theoretically or empirically through the King plot technique~\cite{Cheal2012,King1984}.

\begin{figure}
\subfloat[HR-RIS $385\text{nm}$ vs.\ CLS]{\includegraphics[width=\columnwidth]{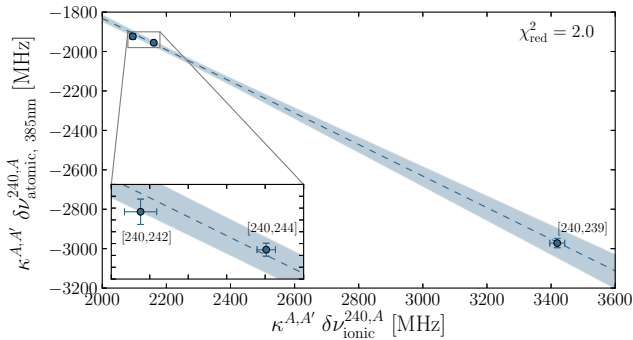}}\\
\subfloat[HR-RIS $388\text{nm}$ vs.\ CLS]{\includegraphics[width=\columnwidth]{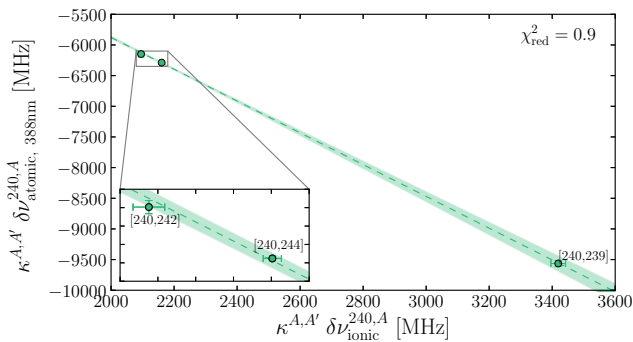}}
\caption{(Colour online) King plots of the (a) $385.210\text{nm}$ and (b) $387.965\text{nm}$ HR-RIS atomic isotope shifts vs.\ the ionic isotope shifts from CLS. The error bars are within the data points. The dashed lines denote the straight lines of best fit. The $68\%$ $(1\sigma)$ confidence bands are included as the shaded areas.}
\label{fig:KP-RIS}
\end{figure}

The King plot allows a direct determination of the atomic mass and field shift factors by examining optical isotope shifts either with respect to changes in mean-squared charge radii obtained from non-optical methods, $\delta \langle r^2 \rangle$, or via a transfer of known atomic factor information from one transition to another provided that each isotope pair has been studied using at least two different transitions. Multiplying equation~\eqref{eq:IS} with a modification factor $\kappa$
\begin{equation}
\kappa^{A,A'} = \frac{m_A m_{A'}}{m_A-m_{A'}}\times\frac{m_{A_\text{ref}}-m_{A'_\text{ref}}}{m_{A_\text{ref}}m_{A'_\text{ref}}},
\end{equation}
removes the dependence of nuclear masses and includes a standard reference pair $A_\text{ref}=244$ and $A'_\text{ref}=240$ for presentation purposes. The modified isotope shifts are then written as
\begin{equation}
\kappa^{A,A'}\delta\nu_{i}^{A',A} = \frac{m_{244}-m_{240}}{m_{244}m_{240}} \times M_{i} + F_{i} K(Z) \kappa^{A,A'}\delta\langle r^2 \rangle^{A',A}\text{,}
\end{equation}
where $i$ (and likewise for $j$) denotes the transition. The modified isotope shifts of two optical transitions $i$ and $j$ may be plotted against each other and should yield a straight line with the atomic factor information contained in the gradient and intercept,
\begin{equation}
\kappa^{A,A'}\delta\nu_{i}^{A',A} = \frac{F_i}{F_j} \kappa^{A,A'}\delta\nu_{j}^{A',A} + \frac{m_{244}-m_{240}}{m_{244}m_{240}} \times \left(M_i - \frac{F_i}{F_j}M_j\right).
\end{equation}
A plot of the atomic isotope shifts determined by the HR-RIS method compared with the ionic shifts from collinear laser spectroscopy is shown in figure~\ref{fig:KP-RIS}. This serves as a consistency check of the measurements using the two techniques. The field shift ratio $F_\text{atomic}/F_\text{ionic}$ is $-0.799(39)$ for the ground state atomic $385.210\text{nm}$ transition and $-2.588(69)$ for the atomic metastable $387.965\text{nm}$ transition. As expected for heavy elements, the $y$-axis intercepts related to the mass shifts $M$ are small, $-232(86)\text{MHz}$ and $-701(151)\text{MHz}$, respectively. 

\begin{figure}
\includegraphics[width=\columnwidth]{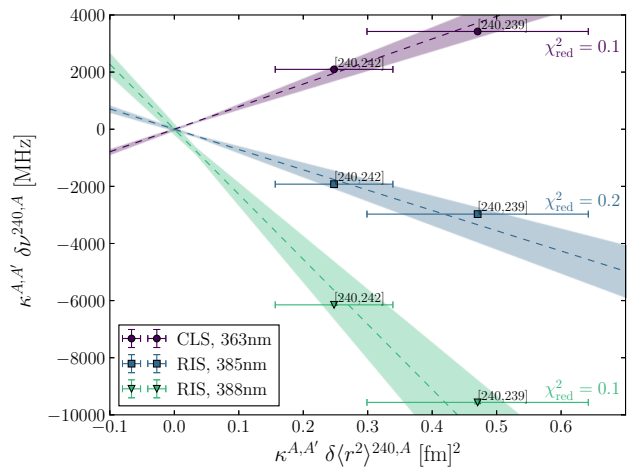}
\caption{(Colour online) King plots of the isotope shifts of all transitions in this work vs.\ changes in mean-squared charge radii evaluated from muonic $X$-ray measurements~\cite{Zumbro1986}. The error bars for $\kappa\times\delta\nu^{240,A}$ lie within the data points. The best fits are shown as the dashed lines with the $68\%$ confidence bands included as the shaded areas.}
\label{fig:KP}
\end{figure}

Absolute charge radii for $^{239,240,242}$Pu have been determined from $X$-ray studies of muonic atoms~\cite{Zumbro1986}. Plotting the modified isotope shifts $\delta\nu^{240,A}$ against the changes in mean-squared charge radii $\delta\langle r^2 \rangle^{240,A}$ allows for a direct evaluation of the atomic factors (see Fig.~\ref{fig:KP} for all transitions in this work). As a fit curve a linear fit through the origin (assuming a negligible mass shift) was used to determine the effective field shift gradients $F_\text{eff} = F \times K(Z)$. The assumption of $M=0$ is based on experimental work in Th~\cite{Sonnenschein2012} where the intercept was consistent with zero. Theoretical work carried out for Fr indicates that the absence of a mass shift contribution causes an uncertainty of $\sim1\%$ in the extracted $\delta\langle r^2 \rangle$~\cite{MartenssonPendrill2000}. The effective field shift for the HR-RIS transitions were determined to be $F_{\text{eff, }385\text{nm}}=-7.1(7)\text{GHz/fm}^2$ and $F_{\text{eff, }388\text{nm}}=-22.8(23)\text{GHz/fm}^2$ from linear fits to the King Plot. The value for the effective field shift factor for the ionic collinear transition was extracted as $F_{\text{eff, }363\text{nm}}=+7.9(6)\text{GHz/fm}^2$. The negative $F$ factors reflect a decrease in $s$-electron density when promoting one electron from either the atomic ground state or metastable state to the excited state, as expected from the atomic configuration. In contrast, the positive $F$ factor in the transition in Pu$^+$ indicates an increase of the electron density at the nucleus agreeing with the assumption of one $f$-electron being transferred to an orbital with lower angular momentum.

\begin{figure}
\includegraphics[width=\columnwidth]{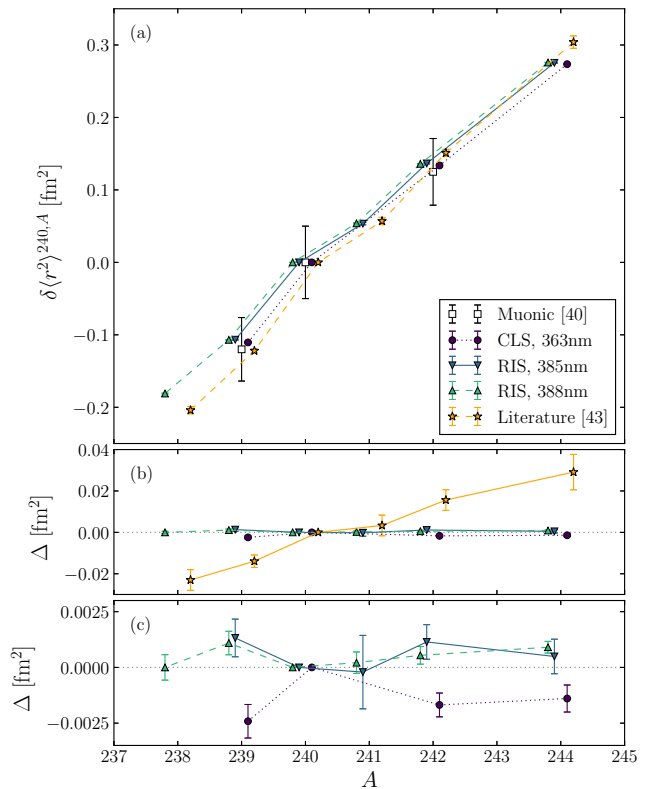}
\caption{(Colour online) (a) Changes in mean-squared charge radii extracted from optical isotope shifts in this work with respect to $^{240}$Pu and compared with literature values from~\cite{Zumbro1986} and~\cite{Angeli2013}. A small horizontal offset was included for display purposes. The error bars for this work lie within the data points. (b) The middle panel highlights the relative change in $\delta\langle r^2\rangle$ for the optical work from Table~\ref{tab:radii} with respect to the average change in mean-squared charge radius for each isotope as obtained from this work; see text for details. (c) In the lower panel, a zoom of the middle panel is provided without~\cite{Angeli2013}.}
\label{fig:radii}
\end{figure}

The extracted changes in mean-squared charge radii are presented in Table~\ref{tab:radii}. Both spectroscopic techniques, collinear laser spectroscopy with fluorescence detection and high-resolution resonance ionisation spectroscopy utilising a narrow linewidth injection-locked laser, provide values for $\delta\langle r^2 \rangle^{240,A}$ with similar sized statistical uncertainties of $\sim5\times10^{-4}\text{fm}^2$. Systematic uncertainties arising from the $F_\text{eff}$ factors are of the order of $10\%$ of $\delta\langle r^2 \rangle^{240,A}$ and therefore dominate any other uncertainties. A graphical comparison is given in Fig.~\ref{fig:radii}. The extracted values for $\delta\langle r^2 \rangle^{240,A}$ from the optical isotope shifts in this work are consistent with those from muonic $X$-ray measurements~\cite{Zumbro1986}, however, have uncertainties approximately one order of magnitude smaller when only comparing statistical uncertainties. A comparison of $\delta\langle r^2 \rangle^{240,A}$ with respect to the average change in mean-squared charge radius per isotope, is provided in the bottom panel of Fig.~\ref{fig:radii}. The relative change in relation to the average is defined as
\begin{equation}
\Delta = \frac 1 N \left(\sum_i^N \delta \langle r^2 \rangle^{240,A}_i\right) - \delta\langle r^2 \rangle^{240,A}_j\;\text{,}
\end{equation}
 where the summation runs over all transitions $i$ studied in this work, $N$ reflects the number of transitions studying a particular isotope and $j$ refers to the transition of interest.

A deviation in $\delta\langle r^2 \rangle^{240,A}$ values compared to~\cite{Angeli2013} might partially be explained by different assumptions of the mass shift constant $M$. In this work, $M=0$ was used for all three investigated transitions. The arc discharge spectra of Pu~\cite{Gerstenkorn1987} were evaluated in~\cite{Angeli2013} against the muonic $X$-ray data from~\cite{Zumbro1986} using $M=+391\text{GHz u}$ assuming an alkali-like transition. For comparison, $\delta\langle r^2 \rangle^{240,A}$ from~\cite{Angeli2013} were plotted against $\Delta$ obtained from the three optical transitions to highlight the influence of the atomic factors on $\delta\langle r^2 \rangle$ as indicated in the middle panel of Fig.~\ref{fig:radii}.

\begin{table*}
\caption{Changes in mean-squared charge radii $\delta\langle r^2 \rangle^{240,A}$ determined from optical work for all investigated transitions. The uncertainties in parentheses are statistical. Contributions from systematic uncertainties on the isotope shifts are denoted by square brackets, contributions arising from the effective field shift factors $F_\text{eff}$ are given in curly brackets (see section~\ref{sec:sys_errors} for details). Literature values are included for comparison~\cite{Zumbro1986,Angeli2013}.}
\label{tab:radii}
\footnotesize
\begin{tabular}{c c  r@{.}l r@{.}l r@{.}l r@{.}l r@{.}l r@{.}l}
\hline\hline
\multicolumn{2}{c}{Transition} & \multicolumn{2}{c}{$F_\text{eff}$}            & \multicolumn{10}{c}{$\delta\langle r^2 \rangle^{240,A}$}\\
\multicolumn{2}{c}{}           & \multicolumn{2}{c}{$[\text{GHz/fm}^2]$}      & \multicolumn{10}{c}{$[\text{fm}^2]$} \\
\multicolumn{4}{c}{}                                                          & \multicolumn{2}{c}{238} & \multicolumn{2}{c}{239} & \multicolumn{2}{c}{241} & \multicolumn{2}{c}{242} & \multicolumn{2}{c}{244} \\
\hline
$385\text{nm}$ & Pu I  & $-$7&1(7)                                            & \multicolumn{2}{c}{}    & $-$0&1067(9)[11]\{105\} & $+$0&0538(14)[11]\{52\} & $+$0&1365(8)[11]\{134\} & $+$0&2754(8)[11]\{271\}\\
$388\text{nm}$ & Pu I  & $-$22&8(22)                                          & $-$0&1810(5)[3]\{174\}  & $-$0&1070(5)[3]\{103\}  & $+$0&0565(4)[3]\{51\}   & $+$0&1359(4)[3]\{131\}  & $+$0&2758(2)[3]\{266\}\\
\hline
$363\text{nm}$ & Pu II & $+$7&9(5)                                            & \multicolumn{2}{c}{}    & $-$0&1105(7)[1]\{69\}   & \multicolumn{2}{c}{}    & $+$0&1337(5)[1]\{84\}   & $+$0&2735(6)[3]\{173\}\\
\hline
\multicolumn{2}{c}{Muonic $X$-rays~\cite{Zumbro1986}} & \multicolumn{2}{c}{}  & \multicolumn{2}{c}{}    & $-$0&120(66)            & \multicolumn{2}{c}{}    & $+$0&125(68)            & \multicolumn{2}{c}{}\\
\multicolumn{2}{c}{Literature~\cite{Angeli2013}}      & \multicolumn{2}{c}{}  & $-$0&204(5)             & $-$0&122(3)             & $+$0&054(5)             & $+$0&151(5)             & $+$0&304(8)\\
\hline\hline 
\end{tabular}
\end{table*}

\section{Systematic Error Budget}
\label{sec:sys_errors}

As the statistical uncertainties on the isotope shifts (and therefore also for the extracted changes in mean-squared charge radii) are comparably small for both the HR-RIS and CLS laser spectroscopy techniques, a further investigation of systematic uncertainties was considered. These are of experimental nature and arise from the conversion of some detected variable to frequency changes. Naturally, the frequency conversion mechanisms differ for both methods but nevertheless can independently introduce uncertainties on the isotope shifts. These will directly affect the extracted $\delta\langle r^2 \rangle$, however, slight changes in the King plots are also expected. For the sake of the arguments presented in this section, the previously mentioned values for the atomic factors will be used.

In addition, the influence of atomic factors is also shortly discussed as those also directly affect the extraction of mean-squared charge radii.

\subsection{Resonance Ionisation Spectroscopy}
\label{sec:sys_errors_RIS}

For the resonance ionisation studies, the master \textsc{Ecdl} laser was stabilised against long-term drifts via a scanning Fabry-P\'erot interferometer to a frequency stabilised HeNe laser. By sequentially changing the lockpoint of the master laser, the frequency of the injection-locked laser was scanned over a $10\text{GHz}$ frequency range in total. The change in frequency of the master/slave laser was then determined to be $\Delta\nu = N \times \text{FSR}_\text{FPI} \times  \lambda_\text{HeNe}/\lambda_\text{master}$ where $N$ is the number of FSRs scanned over, a maximum of $41$ for the $10\text{GHz}$ scan range. Combining this with the uncertainty of the FSR, this results in a negligible effect of $0.1\text{MHz}$ on the frequency scale.

The scanning of the laser and the setpoint interval of $\pm5\text{MHz}$ (thus $\pm10\text{MHz}$ for frequency-doubled laser light) has significantly more influence. Due to the data acquisition mode of recording data whenever the master laser frequency had reached the set interval point, an offset in resonance centroids is observed between the two scanning directions of the master laser. This effect has been corrected for by summing the individual spectra together before analysis, however, bias effects may still be present. A systematic uncertainty of $\sim8\text{MHz}$ is therefore attributed to the isotope shifts. 

Accounting for a slight asymmetry in the HR-RIS resonances in the fitting process by implementing an asymmetric Lorentzian contribution in the Voigt profile, the effect on the centroids of the hyperfine structures is of the order of $0.5\text{MHz}$. In total, the isotope shifts are affected by less than $1\text{MHz}$ being much smaller compared to the contribution from other uncertainties.

Only the frequency of the \textsc{Ecdl} laser was determined and recorded. In principle, the injection-locked laser should be lasing at a frequency very close to that, however effects such as a non-optimal lock, cavity mode-pulling and frequency chirps caused by the pump laser pulse may introduce inaccuracies~\cite{Hannemann2007,Hori2009}. The chirp effect, however, is a constant offset on the frequency axis and therefore valid for all isotopes. Any effect on the isotope shift is negligible.

\subsection{Collinear Laser Spectroscopy}
\label{sec:sys_errors_CLS}

Systematic uncertainties to the measurement of isotope shifts may be introduced from the conversion of scanning voltages to frequencies and may be evaluated using~\cite{Mueller1983}
\begin{widetext}
\begin{eqnarray}
\Delta_\text{sys}\left(\delta\nu^{240,A}\right) &=& \nu_\text{laser}\sqrt{\frac{e V_\text{RFQ}}{2 m_{240} c^2}} \left[\frac 1 2 \left(\frac{\delta V_\text{LCR}}{V_\text{RFQ}}+\frac{\delta m}{m_{240}}\right)\underbrace{\frac{\Delta V_\text{RFQ}}{V_\text{RFQ}}}+\frac{\delta V_\text{LCR}}{V_\text{RFQ}}\underbrace{\frac{\Delta\delta V_\text{LCR}}{\delta V_\text{LCR}}}+\frac{\Delta m_{240} + \Delta m_A}{m_{240}} \right] \\
&=& \nu_\text{laser}\sqrt{\frac{e V_\text{RFQ}}{2 m_{240} c^2}} \left[\frac 1 2 \left(\frac{\delta V_\text{LCR}}{V_\text{RFQ}}+\frac{\delta m}{m_{240}}\right)\times 10^{-3}+\frac{\delta V_\text{LCR}}{V_\text{RFQ}}\times 10^{-4}+\frac{\Delta m_{240} + \Delta m_A}{m_{240}} \right],
\end{eqnarray}
\end{widetext}
with $V_\text{RFQ}$ being the bias voltage of the RFQ, $\delta V_\text{LCR}$ the difference in post-acceleration voltage of the light collection region for $A=240$ and $A=A'$ when on resonance, $\delta m = \left\vert m_A - m_{240} \right\vert$ with all masses being the ionic masses and their uncertainties $\Delta m$ according to Table~\ref{tab:ion_masses}. According to~\cite{Charlwood2009} and~\cite{Campbell2002}, $\Delta V_\text{RFQ}/V_\text{RFQ} = 10^{-3}$ and $\Delta\delta V_\text{LCR}/\delta V_\text{LCR} = 10^{-4}$, respectively. The bias of the RFQ is read out on a scan-by-scan basis via a $1:10^4$ resistor stack~\cite{Campbell2002} and its weighted mean is used. In order to obtain $\delta V_\text{LCR} = \left \vert V_\text{LCR}^{240} - V_\text{LCR}^A \right\vert$, the (hyperfine) spectra were fitted as a function of post-acceleration voltage after calibration of the scanning power supply. The absolute errors on $\delta V_\text{LCR}$ arising from the linear calibration fit were typically $<0.1 \text{V}$ and therefore consistent with~\cite{Campbell2002}. The calculated systematic uncertainties for the isotope shifts are consistent with zero for the reference isotope $^{240}$Pu and increase with increasing or decreasing neutron number, $\Delta N$. The systematic uncertainties obtained for the isotope shifts are transferred to the changes in mean-squared charge radii and have been included in Table~\ref{tab:radii}.

\subsection{Influence of the atomic factors}

Atomic factors play an important role in the extraction of changes in mean-squared charge radii such that slight changes in $F$ and/or $M$ can influence $\delta\langle r^2 \rangle$ dramatically. The effective field shift factors as extracted from the slopes of the King plots in Fig.~\ref{fig:KP} possess an uncertainty of approximately $10\%$, stemming from the large uncertainty of the muonic $X$-ray data~\cite{Zumbro1986}. As such, a systematic uncertainty of the order of $10\%$ is introduced on the values of $\delta\langle r^2 \rangle$ in Table~\ref{tab:radii}.

In the extraction of $\delta\langle r^2 \rangle^{240,A}$, a zero mass shift contribution was assumed. This leads to the assumption of the specific mass shift constant $S$ being of identical value but opposite in sign to the normal mass shift constant $N$. In the absence of alkali-like transitions and theoretical work on this complex system, no predictions for $S$ are available; $N$, however, may be calculated using $N = \nu m_e/m_u$ where $\nu$ corresponds to the frequency of the transition, $m_e$ to the mass of the electron and $m_u$ to the atomic mass unit. Incorporating $M=N$ and thus $S=0$ as a fixed intercept into the King plot (as done in~\cite{Angeli2013}) has no influence on the extracted $F_\text{eff}$ values on the precision quoted in Table~\ref{tab:radii}. The difference in the $\delta\langle r^2 \rangle^{240,A}$ compared to the values presented in Table~\ref{tab:radii} is less than two percent.

\subsection{Inclusion of Systematic Uncertainties}

As the experimental effects discussed in this section are correlated, the total systematic uncertainty to be attributed to the isotope shifts is taken as a direct sum of the individual values. In case of HR-RIS this amounts to $\delta\nu_\text{sys}^\text{HR-RIS} = 8\text{MHz}$ and in case of CLS to $\delta\nu_\text{sys}^\text{CLS} \leq 2.5\text{MHz}$. Ultimately, this yields an additional error to $\delta\langle r^2 \rangle _\text{sys,exp}^{\text{HR-RIS}, 385\text{nm}} = 0.0011\text{fm}^2$ and $\delta\langle r^2 \rangle _\text{sys,exp}^{\text{HR-RIS}, 388\text{nm}} = 0.0003\text{fm}^2$ for the high-resolution resonance ionisation spectroscopy measurements as a result of the experimental technique. Similarly, a systematic error of $\delta\langle r^2 \rangle _\text{sys,exp}^\text{CLS} \approx 0.0004\text{fm}^2$ is attributed to the results from the collinear laser spectroscopy investigations. Relative systematic uncertainties originating from the experimental methods are $<0.5\%$ for $\delta\langle r^2 \rangle^{\text{HR-RIS}, 385\text{nm}}$ and $\delta\langle r^2 \rangle^\text{CLS}$, whereas they are of the order of $2\%$ for $\delta\langle r^2 \rangle^{\text{HR-RIS}, 388\text{nm}}$. The difference emerges from the different $F_\text{eff}$ of the three transitions.

The relative uncertainties of the effective field shift factors are of the order of $10\%$, translating to systematic uncertainties $\delta\langle r^2 \rangle _\text{sys,theo}$ of up to $0.03\text{fm}^2$ on the changes in mean-squared charge radii for the isotopes investigated. Taking such uncertainties in $F_\text{eff}$ into account leads to the conclusion that the two experimental methods are in agreement with one another.

\section{Conclusions}
\label{sec:conc}

Long-lived Pu isotopes have been studied using two complementary laser spectroscopic methods, resonance ionisation spectroscopy using the Mainz atomic beam unit and collinear laser spectroscopy at the \textsc{Igisol} facility of the University of Jyv\"askyl\"a. The measurements using HR-RIS included the use of an injection-locked pulsed Ti:Sapphire laser with an intrinsic linewidth of $\sim13\text{MHz}$ affording a direct comparison of the two techniques.

Isotope shifts have been measured on the ground state $5f^67s^2\ ^7F_0 \rightarrow 5f^56d^27s\ (J=1)$ and metastable state $5f^67s^2\ ^7F_1 \rightarrow 5f^67s7p\ (J=2)$ atomic transitions using the HR-RIS method and the hyperfine factors have been extracted for the odd mass nuclei $^{239,241}$Pu. Collinear laser spectroscopy was performed on the $5f^67s\ ^8F_{1/2} \rightarrow J=1/2\; (27523.61\text{cm}^{-1})$ ionic transition with the hyperfine $A$ factors measured for $^{239}$Pu. The King plot method was used to perform a consistency check of the two techniques as well as providing an empirical extraction of the field shift factors for all three optical transitions. Changes in mean-squared charge radii are consistent with those determined by non-optical muonic $X$-ray studies, however, have a precision approximately one order of magnitude greater when only comparing statistical uncertainties. 

A thorough analysis of experimental systematic uncertainties has been performed. Unforeseen systematic errors in either the wavelength determination, method of data acquisition or possible perturbations to the excited state caused by the high laser power used in the ionisation step may account for any discrepancy in absolute value between the two techniques. Within the dominating uncertainty of $\sim10\%$ on $\delta \langle r^2 \rangle^{240,A}$ due to the effective field shift factors the changes in mean-squared charge radii extracted from the different transitions and techniques are consistent.

This work will hopefully stimulate future theoretical efforts in calculating the field shift and mass shift factors for actinide elements, where especially the specific mass shift constant is of importance. Such calculations would provide invaluable input for the current King plots. Furthermore, this work would benefit from additional measurements of absolute charge radii, e.g.\ through electronic $K$- or muonic $X$-ray isotope shifts, in order to provide constraints for future calculations. In addition, it is of high interest to probe the isotope shifts of a single transition which can be accessed by both experimental techniques.

To date, Pu is now the heaviest element studied using the collinear laser spectroscopic technique. In the immediate future, efforts are under way to expand the high resolution studies to other actinide elements, notably thorium and uranium.

\begin{acknowledgments}
We thank P.~Th\"{o}rle-Pospiech and J.~Runke for preparing the Pu filaments. This project has received funding from the European Union's Horizon 2020 research and innovation programme under grant agreement No.\ 654002, the Academy of Finland under the Finnish Centre of Excellence Programme 2012--2017 (Project No.\ 251353, Nuclear and Accelerator-Based Physics Research at \textsc{Jyfl}), the Sciences and Technology Facilities Council (\textsc{Stfc}) of the United Kingdom, the \textsc{Fwo}-Vlaanderen (Belgium), \textsc{Goa/2010/010} (\textsc{Bof} KU Leuven), the \textsc{Iap} Belgian Science Policy (\textsc{BriX} network P7/12) and a Grant from the European Research Council (\textsc{Erc-2011-Adg-291561-Helios}).
\end{acknowledgments}

\bibliographystyle{apsrev}
\bibliography{paper}

\end{document}